# Functional Augmented State Transfer (FAST) Architecture for Computationally Intensive Network Applications

GOPI KRISHNA SUVANAM[1]

**ABSTRACT**

We describe a novel architecture that combines the simplicity of RESTful architecture with the power of functional programming for delivering web-services. Although, RESTful architecture has been quite useful in simplifying the development of scalable systems, it is not suited for all types of network applications. Our architecture improves upon the RESTful architecture to provide scalable framework for computationally intensive network applications. The proposed architecture is ideal for applications that involve data management and data analysis/calculations on data. Data analytics and financial calculations are two areas where the architecture can be applied efficiently.

**Keywords:**  Web Services, RESTful architecture, functional programming, network applications

## 1 RESTFUL ARCHITECTURE

RESTful architecture is an architectural style for building distributed applications using hypermedia. RESTful architecture for network applications has been quite popular because of its simplicity and completeness for use in many web applications [1]. As a result, RESTful services have gained an upper hand over SOAP architecture [2]. Especially with the rapid growth in mobile and cloud computing, RESTful web services have gained more popularity [3].

The power of RESTful architecture comes from the ability to decouple the client and the server. The client can access the services without learning the API in advance. This ensures scalable development. The client can "browse" through the API of a server and discover the functionality in runtime as opposed to knowing the API documentation in advance. This feature of RESTful architecture is known as HATEOAS (Hypermedia As The Engine Of Application State) [4].

## 2 GAP IN RESTFUL ARCHITECTURE

Although there are several great positives of RESTful architecture, strict adherence to RESTful architecture may not be feasible all the time. Several online tools have been flexible with the architecture and some have even altogether moved to a novel architecture. A classic example would be Facebook moving away from RESTful architecture to a Graph architecture [2] [5]. Yahoo also uses a query language (YQL)[3] for delivering its services. Although it has a REST based API for social data, the API does not completely adhere to RESTful architectural principals.

This tweaking of RESTful architecture is because of an inherent gap in the architecture. RESTful architecture has developed in conjunction with the development of world-wide web (WWW) and http protocol. This architecture has been developed to mimic World Wide Web [6]. Simplicity, visibility and scalability of modern web has inspired this. WWW itself was developed to serve static documents, although over time it has become a much richer platform. Nevertheless the congenital bias towards documents is still embedded in RESTful architecture. To sum up Roy Fielding – "The key abstraction of information in REST is a resource. Any information that can be named can be a resource: a document or image, a temporal service (e.g. "today's weather in Los Angeles"), a collection of other resources, a non-virtual object (e.g. a person), and so on" [7]. Resource-representation architecture has a fundamental drawback for certain type of applications. As every piece of information in this architecture has to be modelled as a resource certain information is not easy to represent. For example if we want to find out weather at let's say Los Angeles (34.05° N, 118.25° W) we could use a URI like /weather/LA or /weather/35.05/118.25. But if we want to fine tune it and find out weather of a particular locality in LA we will be forced to make the URI something like /weather/35.050456/118.25090. Thus the number of URIs can theoretically be infinite as the longitude and latitude are continuous variables. There could be practical work around for using a RESTful architecture even in such cases but they fundamentally violate philosophy of RESTful representation. As an example one could do:

Step 1: Post /mylocation

'lat'=35.050456,'long'=118.25090

---

Step 2: Get /mylocation/weather

But we believe this is in fact a hack and not in line with the core philosophy of de-coupled interactions. An easier architecture would be to express weather as a function and to obtain weather of a particular location one can simply call: get_weather(34. 050456, 118. 25090). This architecture draws inspiration from functional programming. If such a function can be exposed as a web-service, in addition to GET/POST and other REST methods, the utility of the service can increase many fold. This might seem something similar to remote procedure call (RPC) in a SAOP protocol, but in effect we are only trying to extend the methods of REST architecture and not violating other principals of the architecture like statelessness etc. For extending this architecture we would like to use concepts from functional programming paradigm.

## 3 Power of Functional Programming

Functional programming is a style of programming where computation is treated as an evaluation of functions instead of evaluation of a series of instructions. Functional programming is a part of declarative programming paradigm where the programmer "declares" what she wants to achieve in the program rather than specifying how to achieve it. The key pillar of functional programming is immutability of state and execution without side effects. To put it shortly: functional programmes avoid changing of state and do not maintain data that is mutable.

There are several advantages of functional programs:

- Being declarative, they eliminate the need for specifying the procedure to do the computation. This reduces the burden of optimising the code.
- Map, reduce, filter and other first-class functions are part of functional programming features. This makes parallelization easy
- Building an application using small well defined functions is easier than building the application using big objects
- As there are no state-changes or side effects, one would get the same results every time one executes a functional call with the same parameters. This makes testing easy.

Functional programming has existed for a very long time. Initially it was thought that functional programming is slower than imperative programming. But as computing power is becoming cheaper this issue is becoming less relevant. The positives of functional programming far outweigh the negatives significantly [8].

## 4 Proposed FAST Architecture

The most useful paradigm in RESTful architecture is the lack of state [9]. This helps in segregating client and the server in a clean fashion, thus supporting:

- Cache optimization
- Eliminating dependency on continuous availability of connection
- Multi-access and queuing

The idea of the proposed architecture is to keep this feature intact and adding functional capabilities on top of this. Thus in the functional augment state transfer (FAST) architecture, the server is split into two components: a REST server for performing regular REST functions and a functional programming server (called lambda server) for performing functional queries. A pure functional server should satisfy the following criteria

1. It should return the same response for the same inputs irrespective of the state of the machine
2. Any two requests sent to the server can be executed in any order with the same results
3. All inputs and outputs are immutable

But pure functional architecture is practically redundant except if it is just as a calculator app. The state of the world changes continuously and the client would also like to maintain some mutable state on the server. For example in weather in LA could change every minute so get_weather(34.05,118.25) will give different results for different times of request. A combination of RESTful server (called REST Machine) and a functional computation engine (called Lambda Machine) could provide a solution.

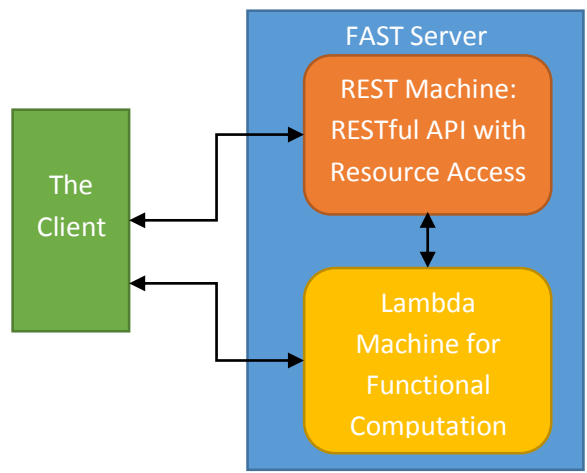

In this framework, the REST machine provides a mechanism to post, update, delete and get resources. Some of these resources could be generated dynamically, in which case the REST API might interact with the Lambda Machine internally. The Lambda Machine also exposes certain functions to the client. The client can request resources or function calls. If it is a function call it is routed to Lambda Machine and if it is a resource requirement it is routed to the REST API

The key ingredients of lambda machine are:

- Any calls to Lambda Machine will not have any side effects
- The data inside lambda machine is immutable. Any mutable data is stored in the REST Server
- User can directly call lambda machine with functions and parameters

Although Lambda Machine might seem similar to remote procedure calls (RPC), it is different in significant ways. Unlike RPC, the proposed lambda machine can interact with a client without the client getting strongly coupled to the server. The calls to Lambda Machine can happen over HTTP. Most importantly the Lambda Machine does not change the state of the system after a call. In other words if the lambda machine is sent two subsequent requests with same data, the result will be same.

FAST API is a combination of the REST API and lambda API. This architecture allows following calls to the FAST API:

- Get or Post a resource (Put and Delete can be included)
- Apply a function on a set of parameters and get the results
- Apply a function on a resource
- Apply a function on a set of parameters and post it to a resource
- Apply a function on the results of another function

A simple way to combine both REST and lambda machines is to interact using a declarative language/template as described in Table 1.

| Query | Comment |
| --- | --- |
| **Table 1: Declarative template for FAST Server** | |
| **Query:**<br>get_weather for latitude=35.05 and longitude =118.25<br>**HTTP style:**<br>GET /lambda/get_weath?latitude=35.05&longitude =118.25 | • You are passing parameters latitude and longitude to the function weather. This request will be picked up by lambda machine<br>• The second format for accessing the same function is through a query language |
| **Query:**<br>Get the value of a book<br>**HTTP style:**<br>Get /fast/pricer/get_value<br>data={"stock_portfolio":"{{ /rest/rest_URI }}"} | • This is an example of mixing regular rest with functional programming<br>• Instead of passing a book as a variable you giving URI of the stock_portfolio which will be "GOT" from the RESTful machine and passed to lambda machine for processing<br>• Notice the URI is embedded inside double curly braces to distinguish it from regular strings |
| **Query:**<br>Get the value of a stock portfolio and store it in another resource<br>**HTTP style:**<br>Post /fast/to_URI<br>'module'='pricer','function'='get_value',<br>data={"stock_portfolio":"{{/rest/rest_URI}}"}<br>Or<br>Post /fast/pricer/get_value<br>data={"stock_portfolio":"{{ /rest/rest_URI }}"}<br>to_uri=<some URI> | • Here you are posting the result obtained from a functional call to a URI<br>• Implementing this would be very tough in regular RESTful architecture<br>• There are two alternatives one is to provide resource to where the data is stored in the URI the other is provide the function in the URI and the resource where data is to posted in the paramters<br>• The second approach is more intuitive |
| **Query:**<br>Apply a list of functions on the same set of data<br>**HTTP style:**<br>Post /fast/pricer<br>fns=['price','delta','gamma','vega'],data={'strike':100,'time':1,'spot':20,'vol':0.2} | • You are passing parameters to four different functions: price, delta, gamma and vega<br>• These functions are part of module "pricer" in the lambda machine<br>• This type of call can help execute multiple functional instructions in a single API query |
| **Query:**<br>Calculate a value and use it within another function<br>**HTTP style:**<br>Get /fast/pricer<br>fns=["price", "delta", "gamma", "vega"],<br>data={"strike":100, "time":1, "spot":20, "vol":"{{/lambda/pricer/implied_vol?strike=100&time=1&spot=20&price=2}})" | • This is a nested function call<br>• You are getting implied vol from some parameters and passing on to another function<br>• Here is where true power of functional programming will come into play<br>• Further levels of nesting will need to be designed more carefully to avoid string conflicts |

Three types of calls can be made to the FAST server:

- Type 1: Functional calls with URI for functions
  - Get /lambda/<lib>/<fun> with parameters
- Type 2: Pure rest calls
  - Get /rest/<path>
  - Post /rest/<path> with {"data":<data>}
  - Post /rest/<path> with {"query":<query>}
- Type 3: Applying function on a resource and Posting functional output to a resource
  - Post /fast/<lib>/<fun> with parameters as URI
  - Post /fast/ with parameters and to_uri

Apart from simple resource and function access one needs higher-order functions for implementing a full functional paradigm. The framework described above can be extended to include such support. Even if support for all higher-order functions are not available, a basic structure for some first-class functions is necessary to perform several computationally intensive operations without hassle.

| Get Map price, delta, gamma, vega from option_pricer on trades | Here trades is a list of parameter sets |
|---|---|
| Reduce add on Map [price] on trades | Map, reduce implementation |

First class functions could be supported as well:

| Get Apply (Apply add on 2) from higher_order_arithmetic on 3 | This should return 5. "add" function from higher_order_arithmetic returns a function |
|---|---|

## 5 ADVANTAGES OF FAST

Apart from the simplicity of RESTful architecture and the power of functional programming, FAST architecture provides the following advantages:

- Clean design: REST calls and Functional computations are segregated. The RESTful server can strictly implement HATEOAS thus avoiding awkward calls with URL parameters or using POST to do calculations etc. The lambda machine can be a pure functional machine with no side effects and immutable state thus making testing, modular development and optimization easy.
- Efficiency: Requests can be automatically parallelized. To parallelize a task one only has to use Map/Reduce/Filter in the request and leave the parallelization to the server.
- Security: As functional computation and resources are segregated, access restrictions can be applied to each of them. Thus a client can access only certain resources and only certain computational functions.
- Ease of development & Scalability: Declarative style of programming lets the client side focus on what it needs for the application rather than how to get it, thus avoiding the need for getting into procedural optimization. With the FAST server one can do multiple steps in a complex computation using a single request.

## 6 APPLICATION

There are several use cases of a FAST architecture. Two areas where it can be applied with efficiency are: financial calculations and analytics. These two use cases are described below.

### 6.1 Financial Calculation App

Financial calculations especially derivative[4] pricing and risk calculations involve complex numerical calculations and Monte Carlo simulations. Performing these calculations on cloud becomes much easier than performing them on local machines. With the increasing use of mobile phones and tables for trading activity, cloud computing is the preferred mode of financial calculations.

Any financial calculation app will have two components: 1. the calculations themselves and 2. Storing and retrieving of financial and user data. First component can be easily served using a lambda machine (functional API server) and the second component can implemented using regular RESTful architecture. Thus a combination of these two i.e. a FAST server is ideal for these kinds of applications.

---

[4] Derivatives are second order financial instruments whose value "derives" from other financial instruments.

### 6.2 Analytics App

Analytics and big data are powerful tools used by all businesses to improve their sales, manage their risks and increase efficiency. Any kind of analytics app will have to manage both the data required for the analytics and also the calculations/analysis. Managing large amount of data and associated computations is not easy using simple service frameworks [10]. Traditional RESTful architecture can easily take care of data related operations. But it fails when it comes to functional calls and requests for some analysis to be performed on the data. This is especially true for real-time analytics. For example if one has to filter out all the users based on certain constraints on age/gender etc. Such a querying system is tough to implement using simple RESTful architecture. Whereas it can be easily implemented using a combination of REST and functional approach. Hence for most analytics apps, FAST architecture is most suitable on the server side.

## 7 A FAST SERVER IMPLEMENTATION IN PYTHON

The sample python code in Table 2 implements a simple FAST server with three types of calls:

| Get Apply <function> from <package> on {'data':<data>} |
|---|
| Get Apply <function> from package on {'uri':<uri_get>} |
| POST Apply <function> from <package> on {'uri':<uri1>} to <uri_post> |

More complex queries cannot be processed using simple url params but can the body of a post request can be used for such calls.

In the server implementation, it is assumed that there is a rest server running separately at "/rest" for getting and posting RESTful resources. The format for sending the first request on http for each of the above calls is:

| Get /lambda/<package>/<function>?data=<data>&to_do=apply |
|---|
| Get /lambda/<package>/<function>?uri=<uri_get>&to_do=apply |
| Post /fast/<package>/<function>?data=<data>&to_do=map&to_uri=<uri><br>data: {<br>"uri":"<uri_get>","to_do":"apply","to_uri":<to_uri>} |

Note: "to_do" parameter can take map, apply, reduce or filter. This is an extension of the basic Lambda Machine provided here to improve the power of the FAST server.

The server is implemented in Python using Flask. Flask is a popular micro web-framework for Python. The code does not concern itself with the implementation of the REST part. A simple REST server is also give below but the server can be implemented any which way, as long as POST and GET methods are defined for the URIs. Pure functional Lambda Machine and interaction between Lambda and RESTful Machines are implemented. Authentication and security controls need to be added on top of this.

Table 2: Python Code for Implementing a Simple FAST Server [5]

```python
# FAST Server

from flask import Flask, request
import json
from importlib import import_module
import restserver

exposed_modules=["operator","basic_arithmetic"]

app = Flask(__name__,template_folder=".")

@app.route('/rest/<path>',methods=["GET","POST"])
def rest_mc(path):
    uri="/rest/"+path
    method = request.method.lower()
    if(method=='get'):
        try:
            return json.dumps(restserver.get_uri(uri))
        except Exception,e:
            return json.dumps({"status":"Server Error"}),500
    if(method=='post'):
        try:
            data=json.loads(request.data) if request.data is not None else request.form

            return json.dumps(restserver.post_uri(uri,data))
        except Exception,e:
            return json.dumps({"status":"Server Error "+str(e)}),500

@app.route('/fast/<module>/<function>',methods=["POST"])
def fast(module,function):
    if(method=='post'):
        args=json.loads(request.data) if request.data is not None else request.form
        restserver.post_uri(args['to_uri'],result)
    else:
        raise NotImplementedError( "This method not implemented" )
@app.route('/lambda/<module>/<function>',methods=["GET","POST"])
def lambda_mc(module,function):
    try:
        ####### Getting the function to evaluate
        if(module not in exposed_modules):
            return json.dumps({"message":"Module not available"}),404
        try:
            mod = import_module(module)
            func=getattr(mod,function)
        except Exception,e:
            return json.dumps({"message":str(e)}),404
        method = request.method.lower()

        ###### Getting the arguments for passing to the function call
        if(method=='get'):
            args = request.args
        elif(method=='post'):
            args=json.loads(request.data) if request.data is not None else request.form

        if 'uri' in args.keys():

            data=restserver.get_uri(args['uri'])
```

---

[5] A version of this code will be shared as an open source project

```python
        elif 'data' in args.keys():
            data=json.loads(args['data'])
        else:
            data={}

        ######## Applying the function on the data
        to_do=args["to_do"]
        if(to_do=="apply"):
            if( isinstance(data,list)):
                result=func(*data)
            if( isinstance(data,dict)):
                result=func(**data)
        elif(to_do=="map"):
            result=map(lambda x: func(*x) if isinstance(x,list) else func(**x),data)
        elif(to_do=="reduce"):
            result=reduce(func,data)
        elif(to_do=="filter"):
            result=filter(lambda x: func(*x) if isinstance(x,list) else func(**x),data)

        ####### Returning the results to the client or posting the results to a URI
        if(method=='get'):
            return json.dumps(result)
        elif(method=='post'):
            return json.dumps(result)
        else:
            raise NotImplementedError( "This method not implemented" )
    except Exception,e:
        return json.dumps({"message":str(e)}),500
if __name__ == '__main__':
    app.run()
    pass
```

```python
#restserver: this implements a simple restserver
data_store={}

######## These functions need to implemented for REST machine to work
def get_uri(uri):
    global data_store
    if(hash(uri) in data_store.keys()):
        return data_store[hash(uri)]
    else:
        return "Resource not found"
def post_uri(uri,data):
    global data_store
    data_store[hash(uri)]=data
    return {"status":"success"}
```

## 8 CONCLUSION

FAST architecture is a combination of RESTful architecture and functional programming paradigm. It is a step forward in implementing efficient network applications including cloud computing applications. There are several areas where such an architecture is much more suitable than RESTful or a SAOP architecture. Such a server has several use cases especially for applications where there is a need for both data management and complex computation. A FAST server can be easily built using existing languages and frameworks.